\documentclass[aps,pra,twocolumn,floats,floatfix,longbibliography,nofootinbib,superscriptaddress]{revtex4-1} 
\pdfoutput=1
\usepackage{amsmath}
\usepackage[dvipsnames]{xcolor}
\usepackage{colortbl}

\usepackage{booktabs}
\usepackage{hyperref}
\usepackage{pgfplots}	\usepackage{pgfplotstable} \usepgfplotslibrary{statistics}
 \pgfplotsset{compat=newest}
 \pgfplotsset{major grid style={dotted,green!50!black}}
\usepackage{tikz}
\usetikzlibrary{plotmarks}

\usepackage{amsopn}

\DeclareMathOperator{\Tr}{Tr}
\newcommand{\unit}[1]{\ensuremath{\, \mathrm{#1}}}
\newcommand{\unitb}{\unit{ms/ \mu m^2}}
\newcommand{\unitd}{\unit{\mu m^2/ms}}

\newcommand{\fig}[1]{Fig.\,\ref{#1}}

\newcommand{\eq}[1]{Eq.\,(\ref{#1})}

\newcommand{\bd}[1]{{\color{black} #1}}

\def \TE {\bd{140 ms}}
\def \TR {\bd{2500 ms}}
\def \FOV {\bd{25.6 cm}}
\def \BW {\bd{2005 Hz/pixel}}
\def \Foufac {\bd{0.75}}
\def \numsl {\bd{14}}
\def \dup {\bd{2.38 \pm 0.02 \unitd}} 
\def \dlo {\bd{2.25 \pm 0.03 \unitd}} 
\def \anglim {\bd{$15^\circ$}} 

\begin{document}

\title{Intra-axonal Diffusivity in Brain White Matter}

\author{Bibek Dhital} 
\email{bibek.dhital@uniklinik-freiburg.de}
\affiliation{Department of Diagnostic Radiology, Medical Physics, University Medical Center Freiburg, Germany}

\author{Marco Reisert} 
\affiliation{Department of Diagnostic Radiology, Medical Physics, University Medical Center Freiburg, Germany}

\author{Elias Kellner}
\affiliation{Department of Diagnostic Radiology, Medical Physics, University Medical Center Freiburg, Germany}

\author{Valerij G. Kiselev} 
\email{valerij.kiselev@uniklinik-freiburg.de}
\affiliation{Department of Diagnostic Radiology, Medical Physics, University Medical Center Freiburg, Germany}

\begin{abstract}

Biophysical modeling is the mediator of evaluating the cellular structure of biological tissues using diffusion-weighted MRI. It is however the bottleneck of microstructural MRI. Beyond the complexity of diffusion, the current development is hindered by the fact that biophysical models heavily rely on diffusion-specific properties of diverse cellular compartments that are still unknown and must be measured {\em in vivo}. Obtaining such parameters by straightforward fitting is hindered by the degenerated landscape of the likelihood functions, in particular, the signal obtained for multiple diffusion directions and moderate diffusion weighting strength is not enough to estimate these parameters: different parameter constellations explain the signal equally well. The aim of this study is to measure the central parameter of white matter models, namely the intra-axonal water diffusivity in the normal human brain. Proper estimation of this parameter is complicated due to (i) the presence of both intra- and extra-axonal water compartments and (ii) the orientation dispersion of axons. Our measurement involves an efficient suppression of extra-axonal space and all cellular processes oriented outside a narrow cone around the principal fiber direction. This is achieved using a planar water mobility filter -- a strong diffusion weighting that suppresses signal from all molecules that are mobile in the plane transverse to the fiber bundle. Following the planar filter, the diffusivity in the remaining compartment is measured using linear and isotropic weighting. We find the specifically averaged intra-axonal diffusivity $D_0 = \dlo $ for the timing of the applied gradients. Extrapolation to the infinite diffusion time gives $D_\infty \approx 2.0\unitd$. This result imposes a strong limitation on the parameter selection for biophysical modeling of diffusion-weighted MRI.
\end{abstract}

\date{\today}

\maketitle

Diffusion-weighted magnetic resonance imaging (dMRI) in brain white matter (WM) has been used to detect tissue anomalies \cite{moseley1990diffusion,werring1999diffusion,kono2001role} and  reconstruct axonal tracts in-vivo\,\cite{basser2000vivo,tuch2004q,tournier2004direct,reisert2011global}. A currently booming research area 
is evaluation of the microstructure of living tissue at the cellular scale much below the nominal MRI resolution. While the role of the light illuminating biological cells is taken by water diffusion as measured by MRI, the role of the microscope is played by biophysical modeling that enables interpretation of diffusion measures in terms of the underlying tissue microstructure. Although much is known about the tissue microstructure from histology, it cannot access parameters that are central to dMRI such as diffusivities inside different cell species. Furthermore, as diffusivities change dramatically upon cell death, they must be measured {\em in\,vivo} -- their physical meaning does not leave much room for alternative (non-MRI) measurement techniques. In this way, the development gets in a vicious circle: Biophysical models need dMRI-specific parameters that should be found using dMRI supplied with biophysical models. 

The problem is exacerbated by the typically featureless shape of commonly acquired dMRI signal. This renders the problem of parameter determination from fitting model to data extremely ill-posed: Essentially different parameter sets can explain the measured data equally well\,\cite{jelescu2016degeneracy}. 

This study is an attempt to break the vicious circle of parameter determination by measuring the intra-axonal water diffusivity in the normal human brain. Neuronal axons are considered as the main contributor to dMRI signal from brain white matter at strong diffusion sensitization\,\cite{veraart2016universal}. While it is hardly possible to completely abandon modeling of dMRI signal, we rely on minimal model assumptions in this study. We developed a dedicated dMRI measurement technique to suppress the signal from extra-axonal space and measure water diffusivity, $D_0$, of the remaining intra-axonal water. 
We find $D_0 = \dlo$, which is rather close to the free water diffusivity at the body temperature ($3\unitd$). This result rules out a large domain of the parameter space available for dMRI signal interpretation.

\begin{figure}[tbp]
   \includegraphics[width=0.49\textwidth, trim = 0pt 0pt 0pt 0pt,clip]{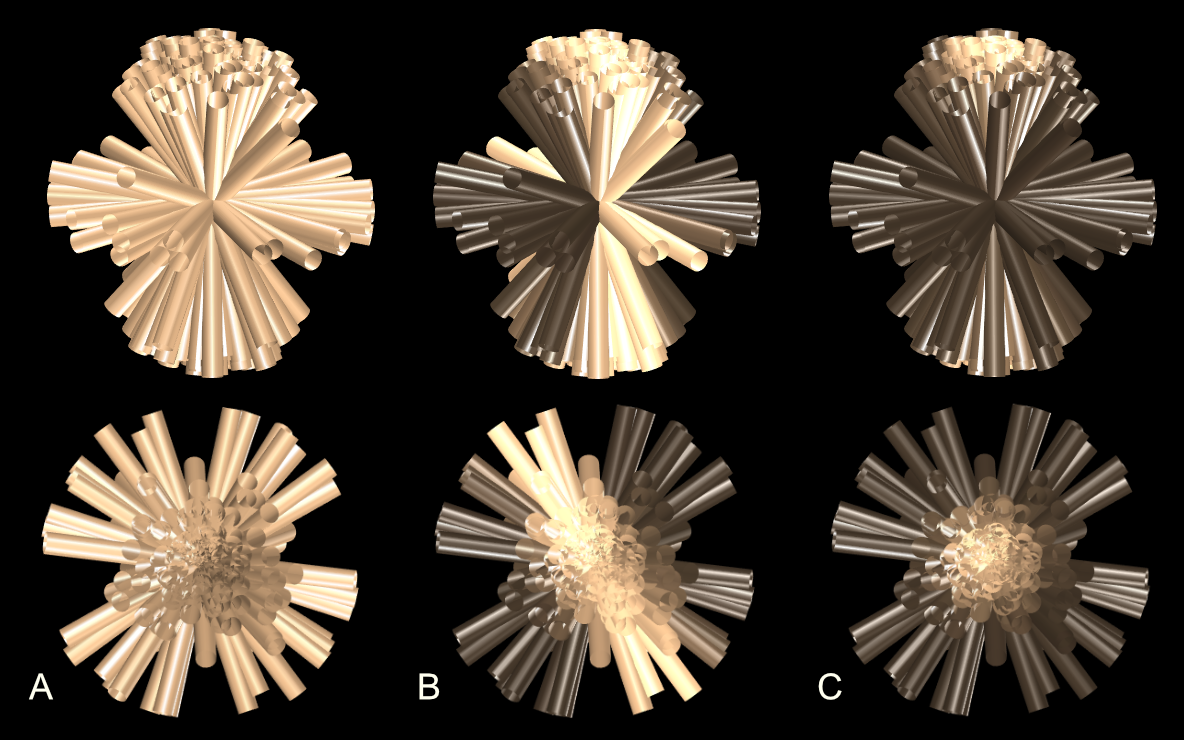}
   \caption{The principle of the measurement.  A: Neuronal axons are represented by cylinders positioned at a common origin to show their orientation distribution (which results in the massive overlapping as a side effect). The in-plane cylinders represent axons incoherent with the principal direction and glial processes in extra-axonal space. The second row shows the top views. B: Application of a linear water mobility filter suppresses extra-axonal space and a part of axons, that are shown with the darker color. Diffusivity measured along the fiber bundle underestimates the true value due to the contribution of tilted axons. C: The planar water mobility filter, which is used in this study, suppresses the signal from all cells, but those that are close to the principal fiber direction. Diffusivity in this direction approaches the ground truth. Isotropic diffusion measurement is applied to eliminate the residual effect of the orientation dispersion. }
    \label{fig:axons}
    \centering 
\end{figure}

\section{Measurement technique matches the target geometry}

dMRI measures the loss of coherence between individual spins, which causes signal attenuation.
The loss of coherence is due to molecular motion in the presence of magnetic field gradients.
In the commonly employed Stejskal-Tanner method\,\cite{stejskal1965spin}, the gradient direction is constant during the measurement so that the resulting signal is sensitized to diffusion in that direction. 
Varying gradient direction during measurement can sensitize the signal to motion in all three spatial direction, which is referred to as the {\em isotropic weighting} or to motion in a selected plane, the {\em planar weighting}. 

For the present measurement, we select the geometry of diffusion weighting to match that of neuronal fiber bundles in white matter, Fig.\,\ref{fig:axons}. 
We apply a strong planar weighting that acts as a {\em planar water mobility filter} by suppressing the signal from water molecules that are mobile in the plane orthogonal to a fiber bundle.
Since the axonal diameter is of the order of a micrometer\,\cite{aboitiz1992fiber}, radial water mobility inside the axons  is negligible. Therefore, the planar filter performs a robust suppression of signal from extra-axonal space, independently of whether water is mobile in the plane or confined in glial processes or axons in the bundle's transverse plane. The suppression would not act on water confined in small compact cells in which water motion is limited in all three directions. However, the presence of such cells in the normal white matter is limited to maximum 2\% as indicated by measurements with strong isotropic diffusion weighting \cite{dhital2017absence,veraart2016universal}.
Following the planar water mobility filter, the remaining signal is primarily contributed by the axons that are close the the axis of the fiber bundle, Fig.\,\ref{fig:axons}.

\begin{figure}
   \includegraphics[width=0.99\columnwidth,clip]{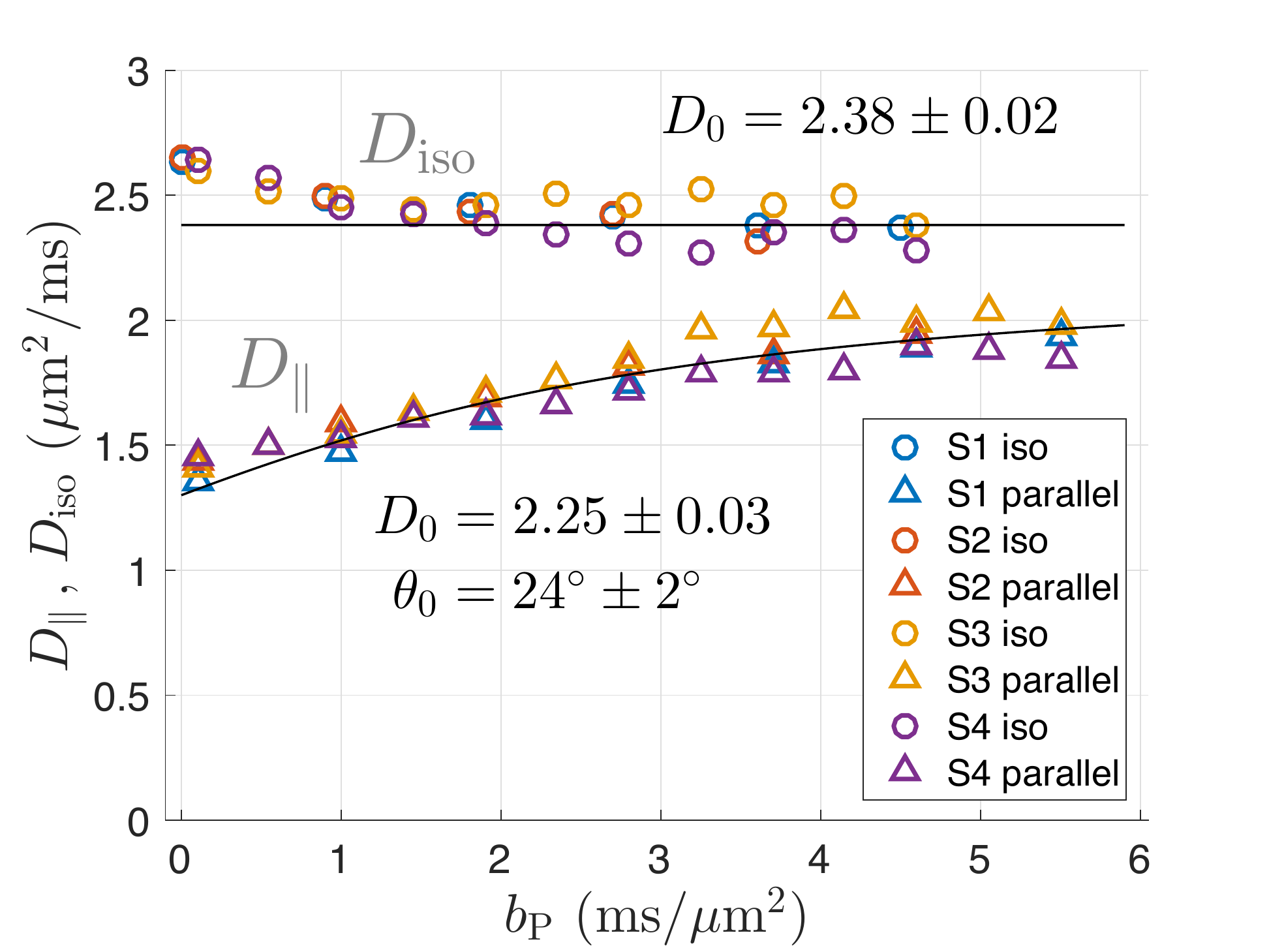}
   \caption{Axial diffusion coefficient, $D_\parallel$, and trace of diffusion tensor, $\Tr D$, in single-fiber voxels in all subjects as functions of the strength of the planar water mobility filter, $b_{\rm P}$.  With increasing $b_{\rm P}$, $\Tr D$ (circles) rapidly approaches the intra-axonal $\Tr D_a$, which is dominated by the intra-axonal diffusivity, $D_0$. $D_\parallel$ (triangles) is measured in the principal fiber bundle direction and underestimates $D_0$ due to the residual axonal orientation dispersion (\fig{fig:axons}). The slow approach to the asymptote contains information about the width of the distribution as discussed in the text. The black lines and the text show the results of fitting the corresponding models to the pooled data for all subjects. }
    \label{fig:diff_all}
    \centering
\end{figure}

The strength of the planar filter is characterized by its b-value, $b_{\rm P}$, which can be understood as the signal suppression by the factor $e^{-b_{\rm P}D}$ in a homogeneous medium with the isotropic diffusivity $D$ (in more detail, the filter b-matrix has the eigenvalues $b_{\rm P}/2\,, b_{\rm P}/2\,,0$). 
The planar weighting reshapes the native axonal orientation distribution by narrowing it around the principal fiber direction. 
Water diffusivity measured along this direction, $D_\parallel$,  approaches the intrinsic intra-axonal diffusivity, $D_0$, but still underestimates it due to the residual dispersion of axons, Figures\,\ref{fig:axons} and \ref{fig:diff_all}. 
It is straightforward (section Methods) to quantify this effect as  
\begin{equation}\label{Dparallel}
D_\parallel = D_0\langle \cos^2\theta \rangle_{\rm F} \,,
\end{equation}
where $\theta$ is the angle an axon makes with the principal fiber direction and the averaging is taken over the filter-reshaped orientation distribution.
The mean $\cos^2\theta$ approaches unity for very strong planar filter, which is obviously limited by the associated decrease in the signal magnitude and hardware limitations.  

The issue of underestimation can be circumvented by following the planar water mobility filter with the isotropic encoding\,\cite{lasic2014microanisotropy,szczepankiewicz2015quantification,dhital2017absence}.
In principle, isotropic encoding measures the trace of diffusion tensor inside axons, but for long diffusion times, the eigenvalues of this tensor can be well approximated by $0\,,0\,,D_0$, so that $\Tr D = D_0$. 
Isotropic encoding gives an estimate of the intra-axonal diffusivity, which is insensitive to the residual axonal orientation distribution.  

The described measurement was performed in the brains of healthy volunteers. Diffusion-weighted signal was measured in many pre-defined directions after application of the planar filter with variable strength in the orthogonal plane (18 directions in subjects 1 and 2 and 30 directions in subjects 3 and 4).  
Data with the weakest filter ($0.1 \unitb$) were used to estimate the unsuppressed local diffusion tensor \cite{basser1994mr}.
The tensor was used to select voxels containing predominantly single fiber bundles (see Methods).
Among these single-fiber voxels, those were further selected in which the principal fiber orientation formed a small angle (less than \anglim) with one of the measurement directions. 
Both $D_\parallel$ and $\Tr D$ were calculated in these voxels as functions of the planar filter strength (\fig{fig:diff_all}) and interpreted in the context of the model presented in Fig.\,\ref{fig:axons} to obtain the intra-axonal diffusivity, $D_0$, and the width of the axonal orientation distribution, $\sigma_0$, (further details in section Methods).

\section{Method}

\subsection*{Experimental Design}
Our measurement technique depends on two ans{\"a}tze
\begin{itemize}
\item Diffusion in the axons is effectively one-dimensional. Since the axon diameter is very small\,\cite{aboitiz1992fiber,caminiti2013diameter}, the radial diffusivity is close to zero for the diffusion time of the order of 100\,ms used in this study. 
\item Diffusion in each compartment is Gaussian: The correlation time of water motion through the cellular environment is much smaller than the diffusion time\,\cite{novikov2010effective}. 
\end{itemize}
In contrast to the majority of the present models, we do not assume the three-dimensional water mobility in the extra-axonal compartment. Our approach copes with possible complex composition of this compartment including one-dimensional cellular processes, restrictions to planes and connected three-dimensional space.

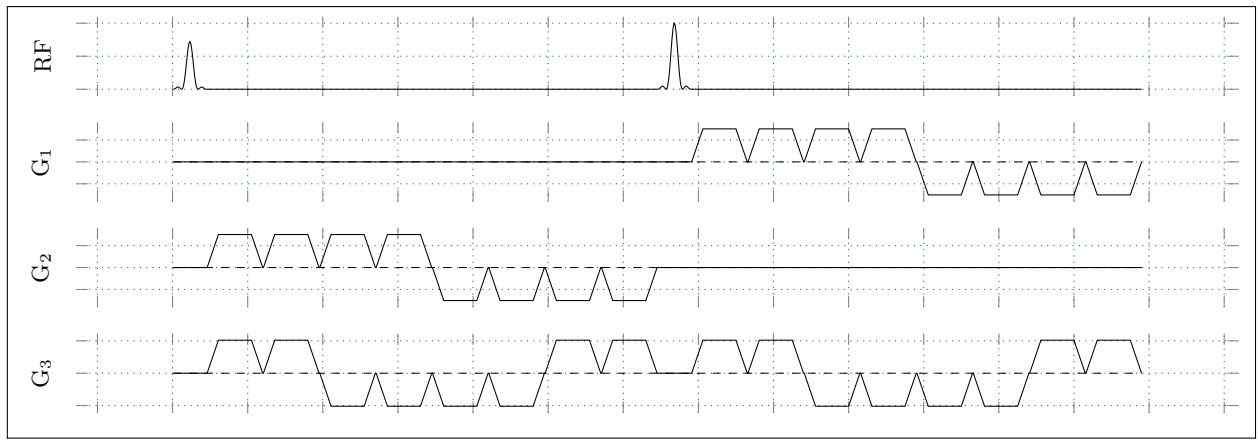
\begin{figure*}[tbhp]
\fbox{
  \begin{tikzpicture}
\begin{axis}[name=plot1,
width=.95\textwidth, height=75pt, grid = major,
xticklabels={,,}, yticklabels={,,}, axis line style={draw=none},
ylabel={RF},
]
\addplot [color=black,]  
  plot[mark= no mark, mark options={fill=white}]
 table[x index=0, y index =1]{Sequence44_2222.dat};
\end{axis}

\begin{axis}[name=plot1,
width=.95\textwidth, height=75pt, grid = major,
xticklabels={,,}, yticklabels={,,},axis line style={draw=none},
xshift=0.0cm,yshift=-40pt, 
ylabel={G$_1$},
]
\addplot [color=black,]
  plot[mark= no mark,] 
 table[x index=0, y index =2]{Sequence44_2222.dat};
  \addplot [] plot[dashed, very thin] table[x index=0, y expr = x*0,]{Sequence44_2222.dat};
\end{axis}

\begin{axis}[name=plot1,
width=.95\textwidth,
height=75pt,
grid = major,
xticklabels={,,}, yticklabels={,,}, axis line style={draw=none},
yshift=-80pt,  
ylabel={G$_2$},
]
\addplot [] plot[] table[x index=0, y index =3]{Sequence44_2222.dat};
\addplot [] plot[dashed, very thin] table[x index=0, y expr = x*0,]{Sequence44_2222.dat};
\end{axis}

\begin{axis}[name=plot1,
width=.95\textwidth,
height=75pt,
grid = major,
xticklabels={,,}, yticklabels={,,},
 ylabel={G$_3$},
yshift=-120pt, axis line style={draw=none},
]
\addplot [] plot[] table[x index=0, y index =4]{Sequence44_2222.dat};
\addplot [] plot[dashed, very thin] table[x index=0, y expr = x*0,]{Sequence44_2222.dat};
\end{axis}
\end{tikzpicture}
}
\caption{Schematic of the sequence used for our measurements. All the three gradient directions are orthogonal to each other and the gradient waveforms do not have any cross terms. The top two gradients provide planar diffusion filtering, the bottom gradient provides an additional linear weighting. Each of these gradient lobes has ramp-up time of 1.5\,ms and flat-top time of 4.3\,ms. Slice selection gradients, spoilers and gradients for imaging are not shown. }
\label{fig:SequenceSchematic}
\end{figure*}

Diffusion-weighted signal for Gaussian diffusion takes the form  
\begin{equation}\label{eq:S=}
S = e^{-\Tr (bD)}
\end{equation}
where $b$ is the b-matrix as defined by applied diffusion-weighting gradients and $D$ is the diffusion tensor.
Both $b$ and $D$ that are real, symmetric matrices.
The signal form is applied to all individual signal-contributing tissue compartments.

Each $b$-matrix can be decomposed into linear $b_{\rm L}$, planar $b_{\rm P}$ and spherical $b_{\rm S}$ components\,\cite{de2016two} in its eigenvector basis, 
\begin{equation}
\mathbf{b} =
b_{\text{L}}\begin{bmatrix}
       0 & 0 & 0  \\
    0 & 0 & 0 \\
   0 & 0 & 1
\end{bmatrix}
+
\frac{b_{\text{P}}}{2}
\begin{bmatrix}
       1 & 0 & 0  \\
    0 & 1 & 0 \\
   0 & 0 & 0
\end{bmatrix}
+
\frac{b_{\text{S}}}{3}
\begin{bmatrix}
       1 & 0 & 0  \\
    0 & 1 & 0 \\
   0 & 0 & 1
\end{bmatrix} \,.
\end{equation}
The diffusion weighting gradient waveforms can be designed to obtain the desired $b$-matrix.
We employed a planar filter to suppress signal from the extra-axonal compartment.
A planar filter is designed by ensuring $b_{\rm L} = b_{\rm S} = 0 $.

The planar filter is designed to take advantage of the difference in the radial diffusivities of the two compartments.
For any diffusion tensor $D$, the signal due to the planar filter can be written as
\begin{align}
\label{eqn:S_planar_tensor}
\ln S =& - \frac{b_{\rm P}}{2}   d_{1} \sin^2 \theta \nonumber \\ 
& -\frac{b_{\rm P}}{2} d_2 \left( 1-  \sin^2 \theta  \sin^2 \phi \right) \nonumber \\ 
& -\frac{b_{\rm P}}{2} d_3 \left( 1- \sin^2 \theta  \cos^2 \phi \right)
\end{align}
where $d_{1,2,3}$ are the largest, middle and the smallest eigenvalues of $D$, $\phi$ is the rotation axis in the $d_{23}$ plane and $\theta$ is the rotation angle.
When the planar filter is normal to the axonal direction, it suppresses the signal only from the extra-axonal compartment.
The filter is therefore appropriate for small $\theta$.
As we increase $b_{\rm P}$, the planar filter increasingly suppresses signals from all compartment that are mobile in the plane leaving only signal from axons that are close to being normal to the plane, \fig{fig:axons}.

In the absence of signal from the extra-axonal compartment, the orientation dispersion of axons impedes direct measurement of intra-axonal diffusivity and the measured apparent diffusion coefficient (ADC) in the parallel direction results in a downward biased estimate.
We circumvent this issue by applying, in addition to the parallel diffusivity, the measurement with the isotropic weighting.
In the absence of any contribution from the extra-axonal compartments, the difference between trace and ADC is only solely due to dispersion.

As shown in Fig.\,\ref{fig:SequenceSchematic}, the planar filtering was obtained by applying two orthogonal gradient waveforms $G_1$ and $G_2$ on either side of the refocusing pulse. 
The resulting b-matrix contains two degenerate eigenvalues that span the plane we intend to filter.
The linear diffusion weighting is obtained by applying a third gradient $G_3$ that is orthogonal to both $G_1$ and $G_2$ (the mutual orthogonality of all three gradients insures the absence of any cross-terms).
Measurements with and without $G_3$ allow us to measure diffusivity of the remaining signal after the planar filter.
To measure the trace of the remaining signal, small portions of $G_1$ and $G_2$ with the b-values equal to that of $G_3$ are re-attributed from the planar filter to the measurement of $\Tr D$.  

\subsection*{dMRI measurements.} In-vivo measurements were performed on \bd{four} informed volunteers in a 3\,T human scanner (Siemens PRISMA, max gradient strength 80\,mT/m, 32 channel receive coil).
The procedure was approved by the ethics board of University Medical Center, Freiburg. 
Written consents were obtained from all volunteers.
For all volunteers, \numsl~ slices were acquired with field of view (FOV) = \FOV, 4\,mm isotropic resolution, echo time (TE) = \TE, repetition time (TR) = \TR, {bandwidth = \BW} and partial Fourier factor of \Foufac.
All imaging parameters, FOV, TE, bandwidth, resolution and number of slices, were kept the same for all measurements.

A simple dMRI sequence comprising of 30 single direction measurements was also measured for two b-values of 0.05 and $1\,\unitb$ to estimate the diffusion tensor\,\cite{basser1994mr}.

In two subjects (S1 and S2), diffusion measurements were performed for \bd{18} directions of $G_3$, with strength of the planar filter \bd{0.1, 1.0, 1.8, 2.7, 3.6 and 4.5\,\unitb}. \bd{To account for signal-to-noise ratio (SNR) loss, data at higher b-values were acquired with more (up to six) repetitions}. In two other subjects, the measurements were performed for \bd{30} directions of $G_3$, and, instead of the repetitions, the $b_{\rm P}$ values were uniformly distributed in the same interval. 
Each planar filter was applied once without the linear gradient, $G_3$, and once with linear gradient with a b-value of 0.45\,\unitb.

\subsection*{Data Analysis.}
 The data was analyzed using in-house written code in Matlab$^{\text{\tiny{\textregistered}}}$.
 All images with corrected for Gibbs-ringing by interpolating the image based on subvoxel-shifts that samples the ringing pattern at the zero-crossings of the oscillating sinc-function\,\cite{kellner2016gibbs}. 
Diffusion tensor \cite{basser1994mr} estimation was performed using the measurement with the weakest planar filter. 
{\em Single bundle voxels} were selected using the fractional anisotropy (FA) along with measures of linearity $ \left( c_l = (\lambda_1-\lambda_2)/\lambda_1 \right)$, planarity $ \left( c_p = (\lambda_2-\lambda_3)/\lambda_1\right)$ and sphericity $ \left( c_s = \lambda_3/\lambda_1 \right)$, where $\lambda_1$, $\lambda_2$, and $\lambda_3$ are the first, second, and third eigenvalues of the diffusion tensor estimated at that voxel.
These coefficients describe proximity of the tensor to a line, plane and sphere\,\,\cite{westin2002processing}.
The single bundle voxels had to fulfill the limit on the fractional anisotropy, $\mbox{FA} > 0.5$ and $c_l\geq0.4$, $c_p\leq0.2$, $c_s\leq0.35$\,\,\cite{fieremans2011white}.
The rest of the analysis focused on single bundle voxels in which the direction of the primary eigenvector reflects the fiber orientation.

The DTI data was used to obtained the relative angle between the primary eigenvector and the measured direction of the gradient waveform $G_3$. Only those single bundle voxels were selected for further analysis in which the principal fiber direction formed an angle less than {\anglim} with one of measured directions. 
For each planar filter, the parallel ADC was estimated by ordinary least squares method where the natural logarithm of the signal were fitted against the two b-values of 0 and 0.45\,\unitb. 
The trace was obtained by a similar fitting, but in this case signal from each planar filter with no linear weighting was fitted with signal from an increased planar filter and a linear weighting of 0.45\,\unitb.
For example, estimating trace after a planar filter of 1.9\unitb, required fitting data obtained from planar filter of 1.9\,\unitb and linear weighting of 0\,\unitb and planar filter of 2.8\,\unitb and linear weighting of 0.45\,\unitb. In other words, a planar filter of 2.8\,\unitb and a linear filter of 0.45\,\unitb can also be considered as a planar filter of 1.9\,\unitb and an isotropic weighting of 1.35\unitb. 
Trace, therefore, could only be calculated up to the second strongest planar filter.

\section{Results}

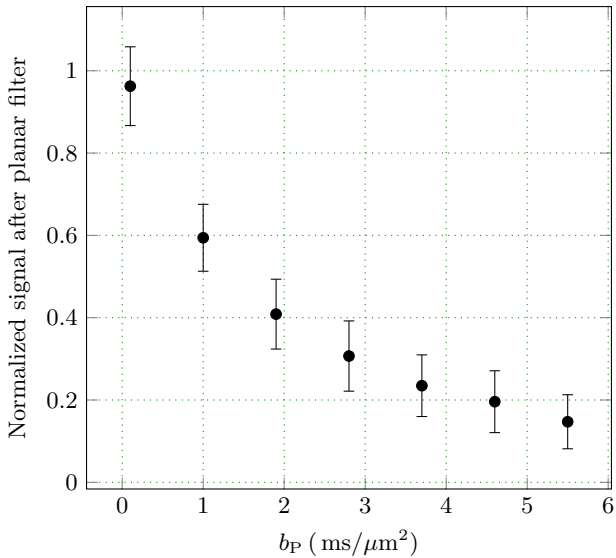
\begin{figure}[tbp]
  \begin{tikzpicture}
\begin{axis}[name=plot1,
width=.99\columnwidth, height=8cm, grid = major,
ylabel={Normalized signal after planar filter},
xlabel={$b_{\rm P}\, (\unitb)$}
]
\addplot [color=black, only marks,]
plot [error bars/.cd, y dir = both, y explicit]
table[x index=0, y index =1, y error index =2]{SNorm_Planar.dat};
\end{axis}
\end{tikzpicture}
\caption{Normalized signal after planar weighting in WM voxels where the relative angle between the fibers and normal to the plane is within \anglim.}
\label{fig:SNorm_Planar}
\end{figure}

An example of the filter effect on the signal is shown in \fig{fig:SNorm_Planar}. 
Figure\,\ref{fig:diff_all} shows the resulted diffusivities obtained from the signal averaged over all selected voxels for both the isotropic and the linear diffusion weighting as a function of planar water mobility filter strength. 
With the increasing filter strength, both results asymptotically level off leaving a small gap in between. 
In what follows, we discuss the limits in more detail. 

The isotropic measurement estimates the trace of the compartment-averaged diffusion tensor, $D$, 
\begin{equation}\label{Diso}
\Tr D = v_F  \Tr D_a + (1-v_F) \Tr D_e \,,
\end{equation}
where $D_a$ and $D_e$ are the intra- and extra-axonal diffusion tensors, respectively, $\Tr D_a \approx D_0$ as discussed above and $v_F$ is the relative fraction of axonal signal after the application of the planar water mobility filter, 
\begin{equation}\label{vF}
v_F = \frac{v_a}{v_a+(1-v_a)e^{-b_{\rm P}D_\perp}} \,,
\end{equation}
where $v_a$ is the genuine water fraction of axons (more precisely, of all effectively one-dimensional processes) and $D_\perp$ is the diffusivity in extra-axonal space in the transverse direction. The value of $v_F$ approaches unity for very strong filter, $b_{\rm P}D_\perp\gg 1$. 

The value of $\Tr D$ for zero filter (\fig{fig:diff_all}) is only slightly above its asymptotic value. 
This implies that the $\Tr D_e$ is only slightly larger than $\Tr D_a$, in agreement with the recent conclusion about a small difference between them \cite{dhital2017absence}. 
We also observe that $\Tr D$ starts leveling off at moderate filter strength. 
This agrees with the expected exponential suppression of extra-axonal space where water can move in the plane normal to the principal fiber direction. 

The axial diffusivity, $D_\parallel$ (\fig{fig:diff_all}) provides two main insights, one from weak filter strengths and the other from strong filter strengths. 
For weak filter, $D_\parallel$ is the weighted mean of intra- and extra-axonal compartments.
While the traces of the two compartments are close, the extra-axonal diffusion tensor has appreciable transverse component. 
Therefore, its longitudinal component is smaller than the intra-axonal diffusivity.
Increasing the filter strength reduces the weight of the extra-axonal contribution and results in some increase in $D_\parallel$. 
For strong filter, the mechanism of the increase in $D_\parallel$ is different.  
It agrees with the argument of narrowing the axonal orientation distribution (\fig{fig:axons}) for the increasing filter strength (as discussed after \eq{Dparallel}). 

We estimated the right-hand side of \eq{Dparallel} by approximating the axonal orientation distribution with a Gaussian distribution of $\sin\theta$ with the variance $\sigma^2$. 
This approximation is justified for distributions effectively narrowed by the planar filter, \fig{fig:axons}. 
This gives 
\begin{equation}\label{Dpar1}
D_\parallel = D_0\left[ \frac{e^{z^2}}{\sqrt{\pi}\, {\rm erfi\,}(z)}-\frac{1}{2z^2}\right] ,\,\,
z = \left(\frac{1}{2\sigma^2} + \frac{b_{\rm P} D_0}{2}\right)^{1/2} \,,
\end{equation}
where the error function of imaginary argument is defined as 
\begin{equation}
{\rm erfi\,}(z) = \frac{2}{\sqrt{\pi}}\int_0^z \!dt\, e^{t^2} \,.    
\end{equation}
For increasing $b_{\rm P}$, $D_\parallel$ slowly approaches its asymptotic value, $D_0$, according to 
\begin{equation}\label{Dpar1asy}
D_\parallel \approx D_0\left[ 1 - \frac{1}{\left(1/\sigma^2+b_{\rm P} D_0\right)^2}\right] \,.
\end{equation}

\begin{figure}
    \centering
    \includegraphics[width=0.99\columnwidth]{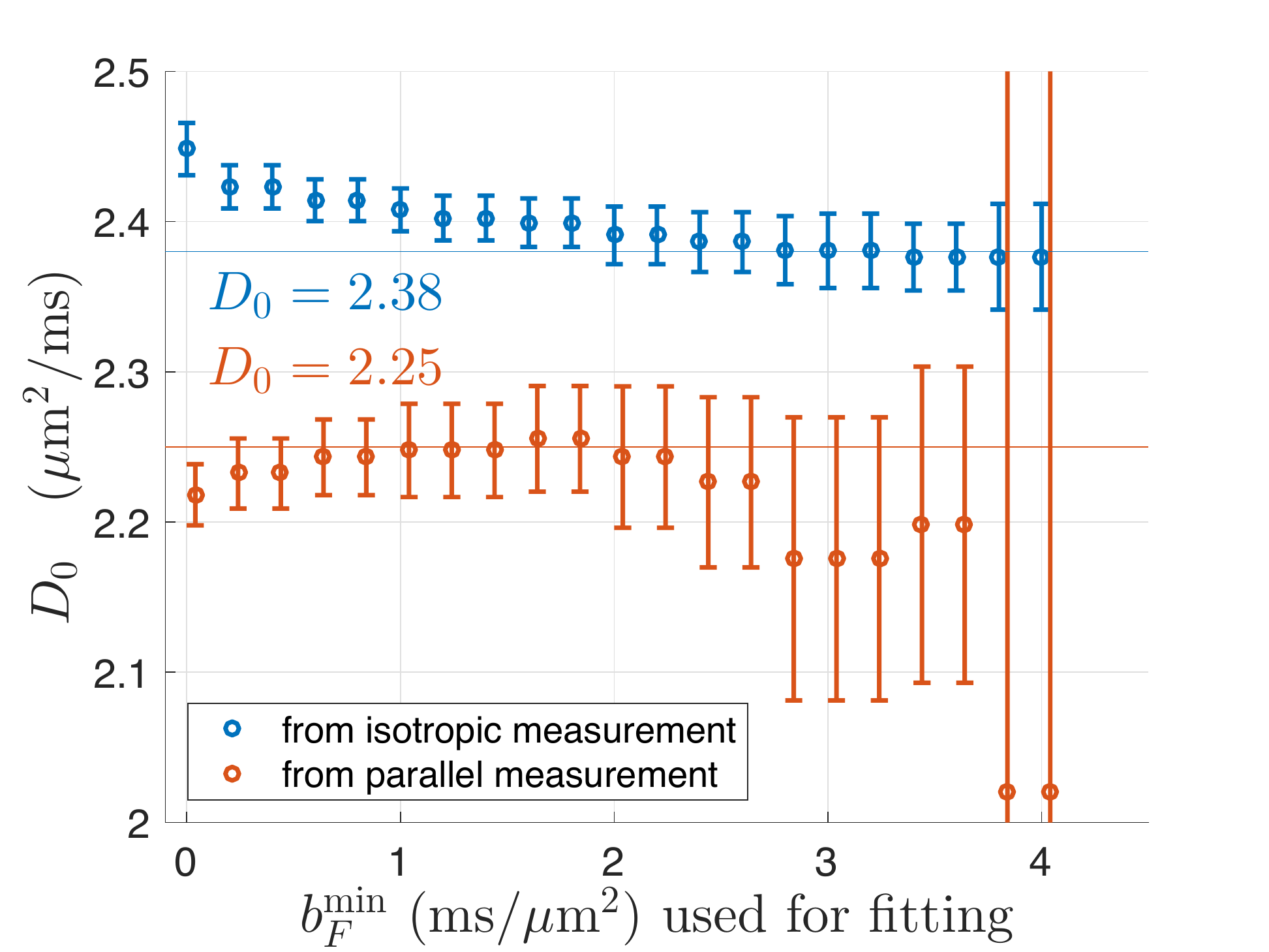}
    \caption{Dependence of the fitted intra-axonal diffusivity on the fitting interval, from $b_{\rm P}^{\rm min}$ upwards. Systematic deviation for small values of $b_{\rm P}$ are due to the extra-axonal contributions not accounted for in the models. The large error for large $b_{\rm P}$ follow from too short fitting intervals. The final selection is shown with the horizontal lines. Note that any other reasonable selection affects the result values within their error bars. }
    \label{fig:fitrange}
\end{figure}

We fitted a constant, $D_0$, which is the asymptotic form of \eq{Diso}, to isotropically measured data and \eq{Dpar1} to corresponding data for the linear weighting. This posed a question about the selection of the data fitting interval, $b_{\rm P}>b_{\rm P}^{\rm min}$. Since both fitting procedures do not take into account the extra-axonal signal (with a number of associated unknown parameters), a too low $b_{\rm P}^{\rm min}$ results in a bias in the estimated parameter. On the other hand, a too large $b_{\rm P}^{\rm min}$ reduces the precision, since fitting is applied within too short intervals, \fig{fig:fitrange}. The choice was made by visual inspection of \fig{fig:fitrange}, some arbitrariness of this action is alike the selection of the commonly used significance level. With this choice made, we obtain the values $D_0=\dup$ and $D_0=\dlo$ for the isotropic measurement ($b_{\rm P}^{\rm min}=3\unitb$) and the single-direction measurement ($b_{\rm P}^{\rm min}=1.2\unitb$), respectively.

The value of $\sigma=0.41\pm 0.03$ and the corresponding angle $\theta=\arcsin \sigma = 27\pm 2^\circ$ overestimate the genuine width of the axonal orientation distribution, $\sigma_0$, due to the additional orientation dispersion introduced by the voxel selection. This effect can be corrected assuming the uniform distribution of voxels within the aperture of $30^\circ$  and the independence of microscopic parameters from the fiber orientation. The corrected values, $\sigma_0=0.41\pm 0.03$ with the corresponding angle, $\theta_0=24\pm 2^\circ$.

\section{Discussion}

The obtained intra-axonal diffusivity of $D_0=\dlo$ is the factor 0.75 smaller than the free water diffusion coefficient at the body temperature. These rather high value resolves the bi-modality of parameter estimation arising from multiple single-direction measurements\,\cite{jelescu2016degeneracy}. The high precision of our result should not be over-interpreted, since it applies to a specific signal averaging; investigation of regional variations was beyond the scope of this study. 

The present result agrees reasonably well with the interval $[1.9,\,2.2]\unitd$ found from $D_a$ obtained at ultra-high $b$, using the $1/b$ scaling of the apparent orientational dispersion in single-fiber populations according to \eq{Dpar1} in the limit when only the intra-axonal signal was present\,\cite{veraart2016universal}. Similar values although with large regional variations were found by studying the combined echo- and diffusion time dependencies\,\cite{veraart2017TE} and using the rotational invariants of signal weighted in multiples single directions and b-values\,\cite{novikov2016mapping}.  

Although not a subjected for direct comparison, data obtained outside the human brain support the present result. Skinner et al.\,\cite{skinner2017rapid} obtained the axial diffusivity of $2.16\pm 0.22\unitd$ in the normal rat spinal cord using the linear water mobility filter (double PFG) in the direction orthogonal to the spinal cord. This should underestimate $D_0$ as illustrated in the middle panel in \fig{fig:axons}. The underestimation can be evaluated using \eq{Dparallel} with the twice reduced $\sigma_0$, which gives $2.16\pm 0.03\unitd$ using $D_0$ and $\sigma_0$ obtained in the present study. This perfect agreement should not be over-interpreted in view of the difference in the investigated tissues. 
Jelescu et al.\,\cite{jelescu2017intra} obtained the axial axonal diffusivity close to $1.7\unitd$ using Gd injection in the rat brain. Since this figure does not account for the axonal dispersion, it should be compared with the reduced diffusivity according to \eq{Dparallel} with the full value of $\sigma_0$. This gives $1.87\pm 0.06\unitd$ indicating a reasonable agreement between the results. 
Jespersen et al.\ investigated fixed pig spinal cord for variable diffusion time and found the long-time intra-axonal diffusivity to be larger than the axial extra-axonal one and close to 1/2 of the free water diffusivity\,\cite{jespersen2017diffusion}. 

The approximately two-fold reduction in the intra-axonal diffusivity relative to the cytoplasm value was also obtained using intra-neurite reporter molecules, NAA, in the rat brain with the reduction factor 0.5\,\cite{kroenke2004nature} and tNAA in the human brain with the reduction factor 0.7\,\cite{ronen2014microstructural}. Equal reduction in diffusivity for different molecules suggests the purely geometric (not chemical) mechanism of this effect. 

The present result for $D_0$ does not agree with the branch selection made for WMTI\,\cite{fieremans2011white} and the fixed values assumed by NODDI\,\cite{zhang2012noddi}. 

The obtained width of the fiber orientation distribution of $\theta_0=24\pm 2^\circ$ agrees reasonably with the values around $19^\circ$ obtained from histology \cite{leergaard2010quantitative,ronen2014microstructural} and MRI-based studies \cite{leergaard2010quantitative,ronen2014microstructural,veraart2016universal}. 

The present method has a few assumptions about the white matter microstructure. Its robustness with respect to deviations from these assumptions and minor corrections to the obtained values are discussed below.  

\subsection*{More about microstructural features}
The majority of white matter models treat extra-axonal space as hindered by axons, but otherwise structureless. 
In reality, white matter is `structurally crowded' being comprised of oligodendrocytes, astrocytes, microglia and vasculature. 
Both astrocytes and oligodendrocytes possess processes. 
The astrocyte processes are smooth, thin and have relatively less branching but extend more than 100 micrometers. The oligodendrocyte processes wrap around the axons to form the myelin layers.
It is quite plausible that glial processes have anisotropic diffusion properties.
The assumption of Gaussian extra-axonal space is based on the effective coarse-graining of structural features for long diffusion times\,\cite{novikov2014revealing,novikov2016quantifying}. 
The course-graining of the whole extra-axonal space within the experimental diffusion time could be effectively hindered by the low permeability of cell membranes\,\cite{nilsson2013noninvasive,yang2017intracellular}.
In such a scenario, the advantage of the present planar filter over the linear filter (\fig{fig:axons}) becomes crucial.

\subsection*{Partial volume effect}
We chose 4\,mm isotropic resolution to increase the SNR, which is crucial when an essential fraction signal is suppressed by application of strong planar filter. However, the question still retains anatomical relevance only when we can clearly resolve not just white and gray matter but also find single bundle WM voxels where the fibers are coherently aligned. 

Owing to high diffusivity in the cerebrospinal fluid, moderate planar water mobility filter efficiently suppresses any partial voluming due to this compartment.
Inclusion of more than one fiber bundle in the selected voxels is limited by the selection of voxels with high anisotropy and by the efficient suppression of possible sub-dominant fiber bundles by the planar filter. 
Again, the difference between the planar and linear filters is crucial for the accurate measurement.
Some admixture of gray matter in the selected voxels is possible, but limited by the selection criteria of high anisotropy. 

\subsection*{Deviations of axonal geometry from ideal cylinders}
The values of $D_0$ from isotropic measurement is larger than that from the linear measurement by the value $0.13\pm 0.04\,\unitd$, which remains to be explained. We speculate that this difference can be attributed to deviations of axon geometry from that of ideally straight cylinders. Such deviations should be effective over the water diffusion length. A simple estimate shows that the curvature with the typical radius $100\,\rm \mu m$ would explain the difference. Assigning such a curvature to all axons does not sound realistic, but a large contribution from a relatively small sub-population cannot be excluded. 
In this context, the value $D_0=\dlo$ obtained with the linear weighting is interpreted as the diffusivity along the axons, while the isotropic weighting adds about $0.07\unitd$ for each transverse direction due to deviations of axons from the ideal cylindrical form. 

\subsection*{Correction for finite diffusion time}
The values of diffusivities found in this study slightly overestimate the genuine long-time values due to the final duration of the applied magnetic field gradients. In more detail, the overestimation results from the finite width of the gradient power spectrum in the following signal form for weak diffusion weighting,
\begin{equation}
    \ln S = -\int \frac{d\omega}{2\pi} |q(\omega)|^2 {\cal D}(\omega) \,,
\end{equation}
where $q(\omega)$ is the Fourier transform of the time-integrated gradient, $g(t)=\gamma G(t)$ and $2{\cal D}(\omega)$ is the autocorrelation function of molecular velocity, which is directly related to the conventionally defined diffusion coefficient\,\cite{novikov2011surface,kiselev2017fundamentals}.  

The magnitude of this effect depends of the form of ${\cal D}(\omega)$ for small $\omega$. We use the theoretical result supported by experimental evidences that this dependence takes the form ${\cal D}(\omega)\approx D_\infty+{\rm Const}\,\omega^{1/2}$\,\cite{novikov2014revealing,fieremans2016invivo}. The calculated correction for the gradients used in our experiments is about $D_0-D_\infty = 0.25\,\rm \mu m^2/ms$. Note that this small correction is only noticeable due to the dependence of ${\cal D}(\omega)$ on the square root of frequency. A linear dependence would result in a negligible difference $D_0-D_\infty$. 

\subsection*{Comparison with other filter techniques}

Our suppression technique is akin to the filter-ex\-change method\,\citep{nilsson2013noninvasive,skinner2015detection}. The filtered-dPFG \citep{skinner2015detection} method uses two pairs of gradient pulses perpendicular to each other \citep{callaghan2002locally} and has been used to show local anisotropy in macroscopically isotropic material including gray matter \citep{komlosh2007detection}. 

Assuming that extra-axonal compartment is an axially symmetric tensor, both the filtered-PFG method and the planar filter would have equivalent suppression efficiency for the extra-axonal compartment.
However, these two methods differ in their signal suppression for the dispersed axons.
While suppression due to the planar filter only depends on the polar angle $\theta$ between the axonal direction and normal to the plane, for filtered-PFG the suppression also depends on the azimuthal angle $\phi$, \eq{eqn:S_planar_tensor}. 
Therefore, the suppression efficiency filtered-PFG is lower for dispersed axons. 
Hence, even in the absence of extra-axonal compartment, parallel ADC obtained after with filtered-PFG method produces a greater downward bias of the true intra-axonal diffusivity than that obtained with the planar filter. 
This probably explains why we obtain a constant increase in parallel ADC with increasing filter weighting but application of filtered-PFG with similar filter weighting showed less effect\,\cite{skinner2017rapid} as discussed above. 

\subsection*{Outlook}
In order to fully realize the clinical potential of diffusion MRI, it is important to understand how the resulted microstructural measures change in response to pathology. This was not however the aim of the present study  that focused on providing an accurate estimate of the intra-axonal water diffusion coefficient. This quantity is central for biophysical modeling that mediates the evaluation of microstructural parameters from the dMRI signal. While creation a solid ground for modeling efforts is the main focus of this study, testing the sensitivity of intra-axonal diffusivity to diverse neurological diseases remains the aim of future work. 

\subsection*{Acknowledgement}
We thank Dmitry S.\ Novikov for useful discussions. 

%

\end{document}